\documentclass[a4paper,12pt]{article}
\usepackage{graphicx}
\begin{document}
\begin{center}
\Large\bf
Anomalous net-baryon-rapidity spectra at RHIC\\[2.6cm]
\large
Georg Wolschin\footnote{E-mail: wolschin@uni-hd.de\hspace{5cm}
http://wolschin.uni-hd.de}\\[.8cm]
\normalsize\sc\rm
Institut f\"ur theoretische Physik der Universit\"at,
D-69120 Heidelberg, Germany\\[.6cm]
(Submitted January 02, 2003)\\[3.0cm]
\end{center}
\bf
\rm
Net-baryon rapidity distributions in central Au+Au collisions at
$\sqrt{s_{NN}}$ = 200 GeV are shown to
consist of three components.
The nonequilibrium contributions are accounted for in a
Relativistic Diffusion Model. Near midrapidity, a third
fraction containing $Z_{eq}\simeq 22$
protons reaches local statistical equilibrium in a
discontinuous transition.
It may be associated with a deconfinement of the
participant partons and thus, serve as a signature
for Quark-Gluon Matter formation.
\newpage
Rapidity distributions of participant (net) baryons are very sensitive
to the dynamical and statistical properties of nucleus-nucleus
collisions at high energies. Recent results for net-proton rapidity spectra in
central Au+Au collisions at the highest RHIC energy
of $\sqrt{s_{NN}}$ = 200 GeV show
an unexpectedly large rapidity density
at midrapidity. The BRAHMS collaboration finds \cite{lee02}
dN/dy = 7.1 $\pm$ 0.7 (stat.) $\pm$ 1.1 (sys.) at y = 0.

The  $\Lambda,\bar \Lambda$ feed-down corrections reduce this
yield by 17.5 per cent \cite{lee02} when performed in
accordance with the PHENIX $\Lambda-$results \cite{adc02} at 130 GeV,
but the amount of stopping remains significant, although
a factor of about 4 smaller as compared to
Pb+Pb at the highest SPS energy. (A corresponding STAR result \cite{adl01}
for y=0 at 130 GeV does not yet include the feed-down correction). Many of
the available numerical microscopic models encounter
difficulties to predict the net-proton yield in the
central midrapidity valley of the distribution,
together with the broad peaks at the
detected positions.

In this Letter I propose to interpret the data in a
nonequilibrium-statistical Relativistic Diffusion Model.
The net baryon rapidity distribution at RHIC energies
emerges from a superposition of the beam-like nonequilibrium
components that are broadened in rapidity space through
diffusion due to soft (hadronic, low $p_{\perp}$) collisions and
particle creations, and a statistical
equilibrium (thermal) component at midrapidity
that arises from hard (partonic, high $p_{\perp}$) processes.

At RHIC energies, the underlying distribution
functions turn out to be fairly well separated in rapidity space.
Since the transverse degrees of freedom are in (or very close to)
thermal equilibrium, they are expected to decouple from the
longitudinal ones. The time evolution of the distribution
functions is then governed by a Fokker-Planck
equation (FPE) in rapidity space
\cite{wol99,wols99,lav02,alb02,ryb02,biy02}. In the more general case of
nonextensive (non-additive) statistics \cite{tsa88} that accounts
for long-range interactions and violations
of Boltzmann's Stosszahlansatz \cite{lav02,alb02,alb00} as well as for
non-Markovian memory (strong coupling) effects \cite{wol02,alb00},
the resulting FPE for the rapidity y in the center-of-mass frame is

\begin{equation}
\frac{\partial}{\partial t}[ R(y,t)]^{\mu}=-\frac{\partial}
{\partial y}\Bigl[J(y)[R(y,t)]^{\mu}\Bigr]+D(t)
\frac{\partial^2}{\partial y^2}[R(y,t)]^{\nu} .
\end{equation}\\
Since the norm of the
rapidity distribution has to be conserved, $\mu$ = 1 is
required, and the
nonextensivity parameter that governs the shape of the
power-law equilibrium distribution becomes q = 2 - $\nu$ \cite{tsa88}.
In statistical equilibrium,
transverse mass spectra and transverse momentum
fluctuations in relativistic systems at SPS-energies
$\sqrt{s_{NN}}$=17.3 GeV require values
of q very slightly above one, typically
q = 1.038 for produced pions in Pb+Pb [10].
For $q \rightarrow1$, the equilibrium distribution
converges to the exponential Boltzmann form,
whereas for larger values of q (with $q < 1.5$) significantly
broader equilibrium distributions are obtained, and
the time evolution towards them becomes superdiffusive \cite{tsa88,wil00}.

To study rapidity distributions in multiparticle systems at
RHIC energies in a nonequilibrium-statistical framework
\cite{wol99,wols99,lav02,alb02,ryb02},
I start with q = $\nu$ = 1 corresponding to the standard FPE.
For a linear drift function
 \begin{equation}
J(y)=(y_{eq}- y)/\tau_{y}
\end{equation}
with the rapidity relaxation time $\tau_{y}$, this is
the so-called Uhlenbeck-Ornstein process, applied to the
relativistic invariant rapidity. The equilibrium value is $y_{eq}=0$
in the center-of-mass for symmetric systems, whereas
$y_{eq}$ is calculated from the given masses and
momenta for asymmetric systems. Using $\delta-$function
initial conditions at the beam rapidities $\pm y_{b}$ ($\pm $5.36
at p=100 GeV/c per nucleon), the equation has analytical Gaussian
solutions. The mean values shift in time towards
the equilibrium value according to
\begin{equation}
<y_{1,2}(t)>=y_{eq}[1-exp(-2t/\tau_{y})] \pm y_{b}\exp{(-t/\tau_{y})}.
\end{equation}
For a constant diffusion coefficient $D_{y}$, the variances of both
distributions have the well-known simple form
\begin{equation}
\sigma_{1,2}^{2}(t)=D_{y}\tau_{y}[1-\exp(-2t/\tau_{y})],
\end{equation}
whereas for a time dependent diffusion coefficient $D_{y}(t)$
that accounts for collective (multiparticle) and memory effects
the analytical expression for the variances becomes more involved \cite{wol02}.
At short times  $t/\tau_{y}<<1$, a statistical description
is of limited validity due to the small number of interactions.
A kinematical cutoff prevents
the diffusion into the unphysical region $|y|>y_{b}$. For larger
values of $t/\tau_{y}$, the system comes closer to statistical
equilibrium such that the FPE is valid.

Since the equation is linear, a superposition of the distribution
functions emerging from $R_{1,2}(y,t=0)=\delta(y\mp y_{b})$
yields the exact solution,
with the normalization given by the total number of net baryons
and the value of $t/\tau_{y}$ at the interaction time t=$\tau_{int}$
(the final time in the integration of (1)) determined by
the peak positions \cite{wol99}.
This approach has also been applied sucessfully to produced
particles at RHIC energies \cite{biy02}, although there the initial conditions
are less straightforward.

The microscopic physics is
contained in the diffusion coefficient. Macroscopically,
the transport coefficients are related
to each other through the
dissipation-fluctuation theorem (Einstein relation)
with the equilibrium temperature T
\begin{equation}
D_{y}=\alpha\cdot T\simeq f(\tau_{y},T).
\end{equation}
In \cite{wol99} I have obtained the analytical result for $D_{y}$
as function of T and $\tau_{y}$ from the condition
that the stationary solution of (1) is equated
with a Gaussian approximation to the thermal equilibrium
distribution in y-space (which is not exactly Gaussian, but
very close to it) as
\begin{equation}
D_{y}(\tau_{y},T)=\frac{1}{2\pi\tau_{y}}\Bigl[c(\sqrt{s},T)m^2T\cdot
(1+2\frac{T}{m}
+2(\frac{T}{m})^2)\Bigr]^{-2}exp(\frac{2m}{T})
\end{equation}
with $c(\sqrt{s},T)$ given in \cite{wols99} in closed form. This allows
to maintain the linearity
of the model and hence, to solve the FPE analytically,
although small corrections are to be expected. They cause
minor deviations in the calculated rapidity distributions that
are within the size of the
error bars of the experimental data at SPS energies.

In the linear model, net baryon rapidity spectra at
low SIS-energies are well reproduced,
whereas at AGS-, SPS- and RHIC energies I find discrepancies to the
data that rise strongly with $\sqrt{s}$.
At SPS energies, this has recently been confirmed
in a numerical calculation \cite{lav02,alb02} based on a nonlinear drift
\begin{equation}
J(y)=-\alpha\cdot m_{\perp}sinh(y)\equiv -\alpha\cdot p_{\parallel}
\end{equation}
with the transverse mass $m_{\perp}=\sqrt{m^2 + p_{\perp}^2}$,
and the longitudinal momentum $p_{\parallel}$.
Together with the dissipation-fluctuation theorem (5), this
yields exactly the Boltzmann distribution as the stationary
solution of (1) for $\nu=q=1$. The corresponding numerical
solution with $\delta-$function initial conditions
at the beam rapidities is, however, only approximately correct
since the superposition principle is not strictly valid
for a nonlinear drift. Still, the numerical result shows
almost the same large discrepancy
between data and theoretical rapidity distribution
as the linear model. In a q=1 framework, the net
proton distribution in Pb+Pb at the highest SPS energy requires
a rapidity width coefficient $\sqrt{D_{y}\tau_{y}}$
that is enhanced beyond the
theoretical value (5) by a factor of
$g(\sqrt{s})\simeq2.6$ due to memory and collective effects
\cite{wol99,wols99,wol02}, Fig.1.

Alternatively, a
transition to nonextensive statistics \cite{tsa88,alb00,wil00}
maintaining the weak-coupling diffusion coefficient
from (5) requires a value of q that
is significantly larger than one. In an approximate
numerical solution of (1) with the nonlinear drift (7),
q=1.25 has been determined for the net-proton
rapidity distribution in Pb+Pb collisions at the SPS \cite{lav02,alb02}.
The only free parameter is q, whereas in the
linear q=1 case the enhancement of $D_{y}$ beyond (5) is
the only parameter.

This value of q in the nonlinear model
is considerably larger than the result
$q(\sqrt{s_{NN}})$=1.12 extrapolated from Wilk et
al. \cite{ryb02} at the SPS-energy
$\sqrt{s_{NN}}$
= 17.3 GeV. Here, the relativistic diffusion
approach is applied
to produced particles in proton-antiproton collisions
in the energy range $\sqrt{s}$ = 53 GeV - 1800 GeV,
and used to predict LHC-results. The nonlinearity $q>1$
appears to be an essential feature of the
 $p\bar p$ data. The larger value
of q in heavy systems as compared to $p\bar p$
at the same NN-center-of-mass energy
emphasizes the increasing superdiffusive
effect of multiparticle
collisions both between participants,
and between participants
and produced particles. It is, however, conceivable
that both a violation of (5) due to memory effects, and $q>1$
have to be considered in a complete description.

Investigating the Au+Au system at RHIC energies in
the linear q=1 model, it is now easy to
show that in central collisions (10 per cent
of the cross-section) the nonequilibrium net-proton rapidity
spectrum calculated with
the standard dissipation-fluctuation theorem
underpredicts the widths of the nonequilibrium
fractions of the experimental distribution \cite{bea02,lee02} significantly.
In this comparison, the temperature T=170 MeV is taken from a thermal fit
of charged antiparticle-to-particle ratios
in the Au+Au system at 200 GeV per nucleon \cite{bea02a,bec01},
and the theoretical value of the rapidity width coefficient 
calculated from the analytical expression (6) is
$\sqrt{D_{y}\tau_{y}}=7.6\cdot 10^{-2}.$
As was shown in \cite{lav02} for SPS results, the discrepancy
persists in case of the
nonlinear drift (7) that yields the exact Boltzmann-Gibbs
equilibrium solution for q=1
\begin{equation}
E\frac{d^{3}N}{d^{3}p}=\frac{d^{3}N}{dy\cdot
m_{\perp}dm_{\perp}d\phi}\propto E\cdot exp(-E/T).
\end{equation}
An enhancement of the weak-coupling
rapidity width coefficient
by a factor of $g(\sqrt{s})\simeq3.7$
due to collective and
memory effects in the system
corresponding to a violation of (5) yields
a good reproduction of the nonequilibrium
contributions with
$\tau_{int}/\tau_{y}$=0.26, but the
midrapidity valley that is present in the data is completely
absent in the extensive nonequilibrium q=1 case, solid curves
in Fig.1 (bottom). This remains true in the
nonextensive case ($1<q<1.5$), with an approximate distribution
function \cite{tsa88,wil00,lav02,alb02,ryb02} that is given by a linear
superposition of Tsallis-like power-law
solutions of (1)
\begin{equation}
R_{1,2}(y,t)=[1-(1-q)\frac{m_{\perp}}{T}
cosh(y-<y_{1,2}(t)>)]^{\frac{1}{1-q}}.
\end{equation}
The dashed curves in Fig.1 show the result for q=1.4,
T=170 MeV and a mean
transverse mass $<m_{\perp}>$=1.2 GeV.
This solution is far from the nonextensive
equilibrium distribution which would be reached for
$<y_{1,2}(t\rightarrow\infty)>=y_{eq}$, and it
is significantly below the midrapidity data. The
result is even worse for larger values of $m_{\perp}$.
In contrast, the Pb+Pb data at SPS energies \cite{app99} are well
described both in the linear model \cite{wol99} (Fig.1, top) and in the
nonlinear case (cf. \cite{lav02} for results with a time-dependent
temperature and an integration over transverse mass).

It turns out, however, that the RHIC data can be
interpreted rather precisely in the linear q=1 framework
with the conjecture that a fraction of
$Z_{eq}\simeq 22$ net protons near midrapidity
reaches local statistical equilibrium
in the longitudinal degrees of freedom. The variance of
the equilibrium distribution $R_{eq}(y)$ at midrapidity is broadened
as compared to the Boltzmann result (dashed curve in Fig.2)
due to collective (multiparticle) effects by
the same factor that enhances the
theoretical weak-coupling diffusion coefficient derived from (5).
This may correspond to a longitudinal expansion
velocity of the locally equilibrated subsystem
as accounted for in hydrodynamical descriptions.
In the nonextensive model, the corresponding equilibrium distribution
is broadened (blue-shifted) according to $q\simeq1.4$.

Microscopically, the baryon transport over 4-5
units of rapidity to the
equilibrated midrapidity region is not only due to
hard processes acting on single valence (di)quarks
that are described by perturbative QCD,
since this yields insufficient stopping. Instead,
additional processes such as the nonperturbative
gluon junction mechanism \cite{ros80} are necessary
to produce the observed central valley. This
may lead to substantial stopping even at LHC energies
where the separation of nonequilibrium and equilibrium
net baryon fractions in rapidity space is expected
to be even better than at RHIC. In the late
thermalization stage \cite{ber01},
nonperturbative approaches to QCD thermodynamics
are expected to be important.

Recent work indicates that one may account
for the observed stopping in heavy-ion collisions at SPS and RHIC
energies with string-model
parameters determined from hadron-hadron collisions \cite{cap02}. If this
was confirmed, the corresponding rapidity distributions would not be
considered to be anomalous from a microscopic point of view.
However, this view does not offer a distinction between
nonequilibrium and equilibrium contributions to the net baryon
rapidity spectra, which both exist at RHIC energies, and
are anomalously broadened.

Macroscopically, the complete solution of (1)
in the q=1 case is a linear superposition
of nonequilibrium and equilibrium distributions (Fig.2, bottom)
\begin{equation}
R(y,t=\tau_{int})=R_{1}(y,\tau_{int})+R_{2}(y,\tau_{int})+R_{eq}(y).
\end{equation}
It yields a good representation of the preliminary
BRAHMS data \cite{lee02}.  (In the $q>1$ case, the corresponding
solution is questionable because
the superposition principle is violated).
Based on (10), the transition from net-proton rapidity
spectra with a central plateau in Pb+Pb at the lower SPS energies \cite{wol02},
via a double-humped
distribution at the maximum SPS energy \cite{app99,wol99,lav02,alb02,wol02}
to the central
valley at RHIC \cite{lee02} is well understood.
It has not yet been possible
to identify a locally equilibrated subsystem of net baryons
at midrapidity below RHIC-energies, although it cannot
be excluded that it exists.
At SPS energies, the data \cite{app99} are well described by the
nonequilibrium distributions, and it is much more difficult
(and probably impossible) to identify a locally equilibrated
component because the relevant rapidity region is comparatively small,
and an equilibrated contribution cannot be separated from the
nonequilibrium components in rapidity space.
In  $p\bar p$-collisions at
$\sqrt{s}=$53-900 GeV,
no convincing signatures of a phase transition were found \cite{gei89}.

Most remarkably, Fig.2 suggests that in central Au+Au collisions
at $\sqrt{s_{NN}}$= 200 GeV
there is no continuous transition from the nonequilibrium
to the equilibrium contribution in net-proton rapidity spectra
as function of time.
This may well be due to a sudden enhancement in the number
of degrees of freedom as encountered in the
deconfinement of participant partons, which enforces a very rapid
local equilibration in a fraction of the system. The central
valley in net-proton rapidity
spectra at RHIC energies could thus
be used as an indicator for partonic processes
that lead to a baryon transfer over more than 4
units of rapidity,
and for quark-gluon plasma formation.

To conclude, I have interpreted recent results for
central Au+Au collisions at RHIC energies in
a Relativistic Diffusion Model (RDM) for
multiparticle interactions based on the interplay of
nonequilibrium and equilibrium ("thermal") solutions.
In the linear version of the model, analytical results for the rapidity
distribution of net protons in central collisions have been
obtained. The anomalous enhancement of the diffusion in rapidity
space as compared to the expectation from the
weak-coupling dissipation-fluctuation theorem due to
high-energy multiparticle effects has been discussed
using extensive and nonextensive statistics.

A significant
fraction of about 14 per cent of the net protons reaches
local statistical equilibrium
in a fast and discontinuous transition which is likely to
indicate parton deconfinement. The precise amount of protons
in equilibrium is related to the experimental
value of the rapidity density close to y=0 and hence, possible changes in the
final data will affect the percentage. It has not yet been
possible to isolate a corresponding fraction of longitudinally
equilibrated net protons in the Pb+Pb system
at SPS energies. Since no signatures of a transition to the
quark-gluon plasma have been observed in $p\bar p$-collisions,
Quark-Matter formation is clearly
a genuine many-body effect occuring
only in heavy systems at sufficiently high energy density.
Consequently, a detailed investigation of
the flat midrapidity valley found at RHIC, and of its energy
dependence is very promising.

\newpage
\Large\bf
Figure captions
\normalsize\rm
\begin{description}
\item[FIG. 1:]
Nonequilibrium contributions to the net-proton rapidity
spectra of Au+Au at $\sqrt{s_{NN}}$ = 200 GeV in
the Relativistic Diffusion Model (RDM) with an equilibrium
temperature of T=170 MeV (bottom). The solid curve is obtained
in the linear model (q=1) with
an enhanced width \cite{wols99,wol02} due to multiparticle effects
according to the preliminary BRAHMS data \cite{lee02}, squares.
The dashed curve corresponds to q=1.4 and $<m_{\perp}>$=1.2 GeV
in the nonlinear model. At SPS energies (top), NA49 data \cite{app99}
for central events (5 per cent) including $\Lambda$ feed-down corrections
are compared with the linear model \cite{wol99,wol02}.
\item[FIG. 2:]
Net-proton rapidity spectra for central collisions of
Au+Au at $\sqrt{s_{NN}}$ = 200 GeV consist
of two nonequilibrium components (solid peaks, top)
plus an equilibrium contribution at T=170 MeV,
dashed curve. It is broadened due to collective
(multiparticle) effects, shaded area, and
after hadronization, it contains
$Z_{eq}\simeq22$ protons. Superposition creates the
flat valley near midrapidity (bottom) in agreement with the
preliminary BRAHMS data points \cite{lee02};
diamonds include $\Lambda$ feed-down corrections at y=0 (17.5 per cent)
and y=2.9 (20 per cent), respectively.
Arrows indicate the beam rapidities $\pm y_{b}$.

\end{description}
\newpage
\vspace{1cm}
\includegraphics[bb=10 80 440 670]{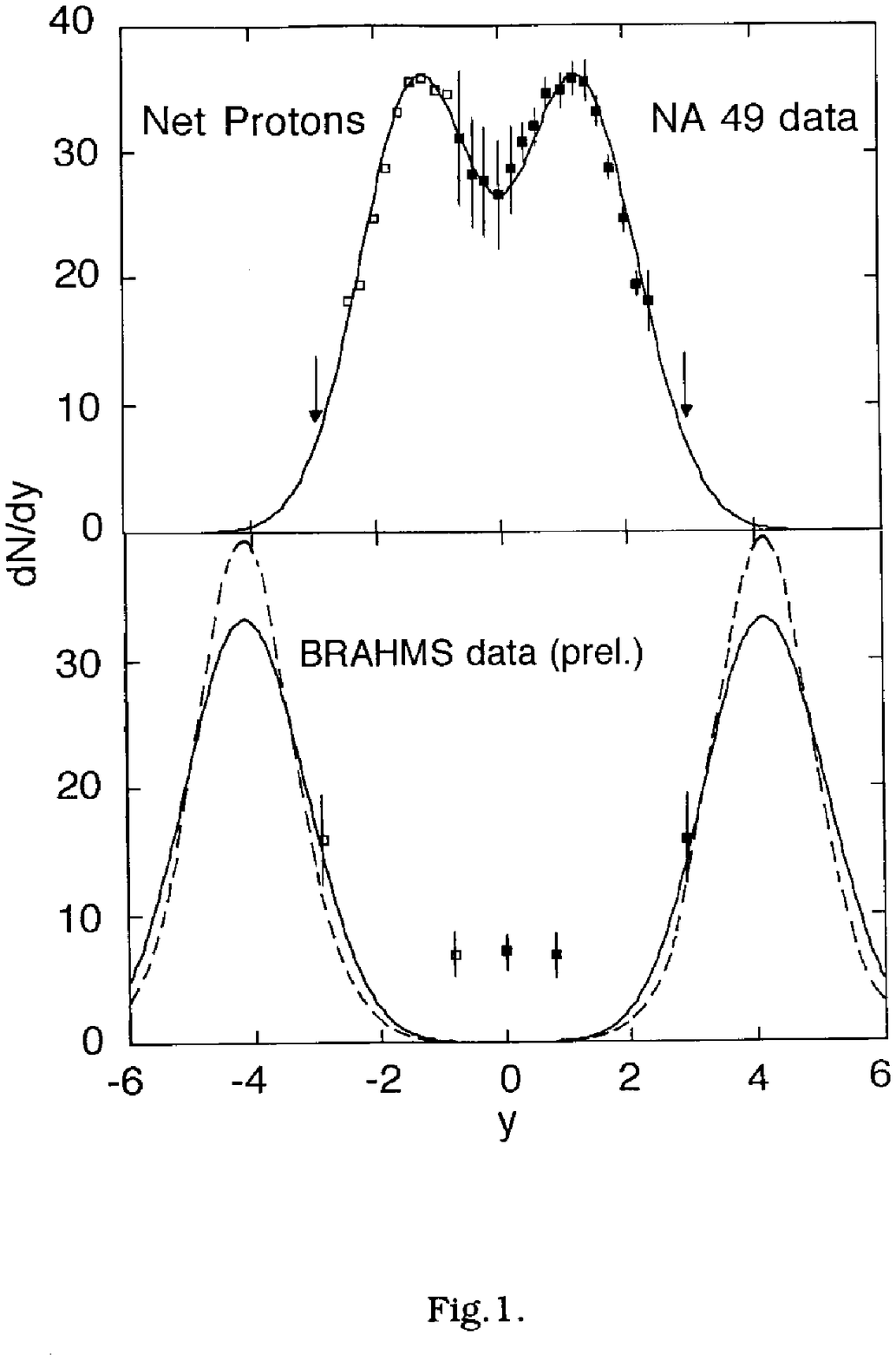}
\newpage
\vspace{1cm}
\includegraphics[bb=10 80 440 670]{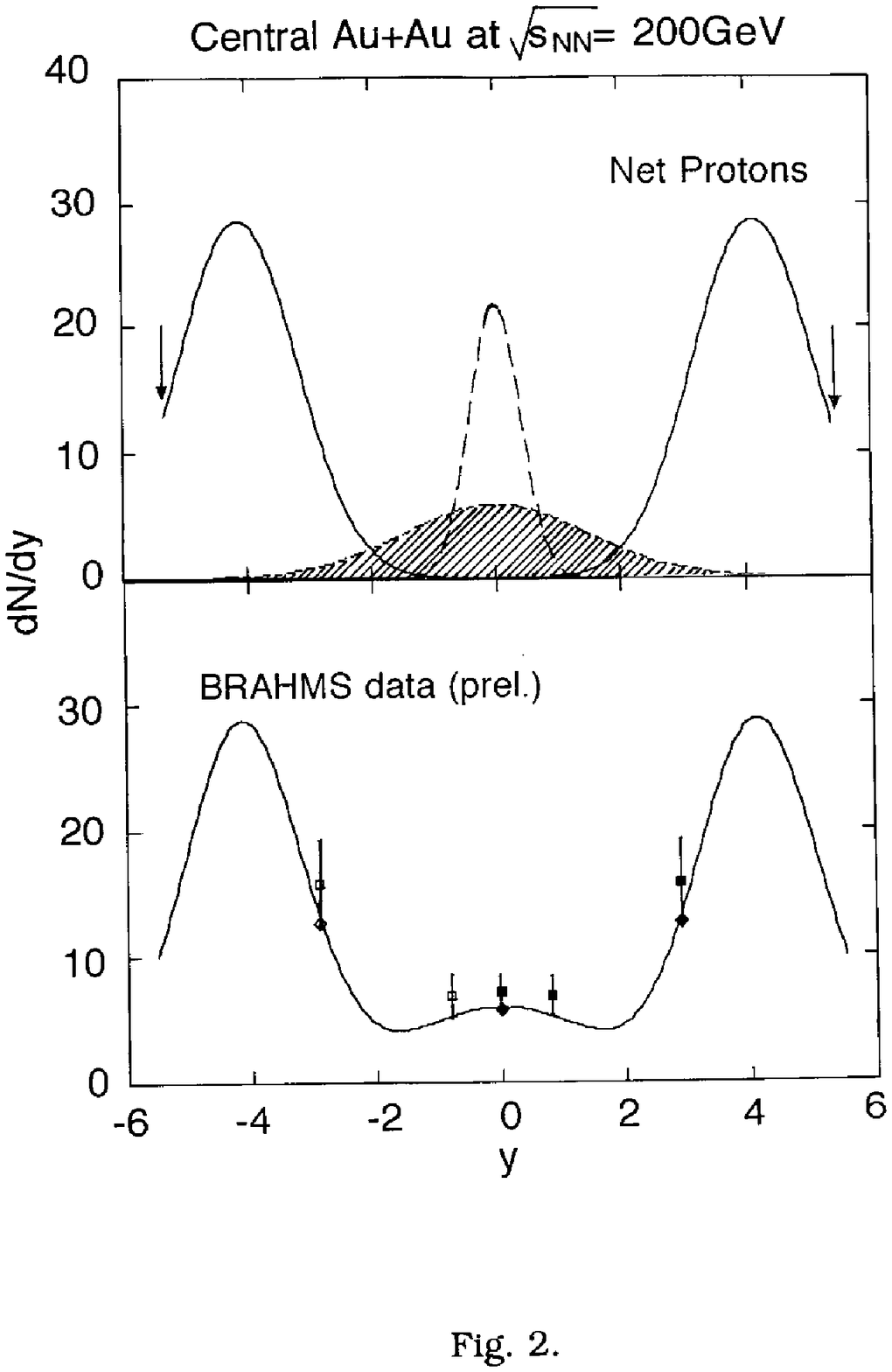}
\end{document}